\documentclass[a4paper,superscriptaddress]{revtex4}
\usepackage{amsmath}
\usepackage{graphicx}
\usepackage[english=american]{csquotes}

\newcommand{\PR}{Phys. Rev.}
\newcommand{\PRL}{Phys. Rev. Lett.}
\newcommand{\PS}{Phys. Scr.}

\begin{document}

\title{Efficient coupling to an optical resonator by exploiting time-reversal symmetry}

\author{Marianne Bader}
\email{marianne.bader@mpl.mpg.de}
\affiliation{Max-Planck-Institute for the Science of Light, G\"unther-Scharowsky-Str. 1 Bldg 24, 91058 Erlangen, Germany}
\affiliation{Institute of Optics, Information and Photonics, University Erlangen-Nuremberg, Staudtstr. 7/B2, 91058 Erlangen, Germany}

\author{Simon Heugel}
\affiliation{Max-Planck-Institute for the Science of Light, G\"unther-Scharowsky-Str. 1 Bldg 24, 91058 Erlangen, Germany}
\affiliation{Institute of Optics, Information and Photonics, University Erlangen-Nuremberg, Staudtstr. 7/B2, 91058 Erlangen, Germany}

\author{Alexander L. Chekhov}
\affiliation{Max-Planck-Institute for the Science of Light, G\"unther-Scharowsky-Str. 1 Bldg 24, 91058 Erlangen, Germany}
\affiliation{Department of Physics, M. V. Lomonosov Moscow State University, Moscow, Leninskie Gory 1 building 62, 119991, Russia}

\author{Markus Sondermann}
\email{markus.sondermann@fau.de}
\affiliation{Max-Planck-Institute for the Science of Light, G\"unther-Scharowsky-Str. 1 Bldg 24, 91058 Erlangen, Germany}
\affiliation{Institute of Optics, Information and Photonics, University Erlangen-Nuremberg, Staudtstr. 7/B2, 91058 Erlangen, Germany}

\author{Gerd Leuchs}
\affiliation{Max-Planck-Institute for the Science of Light, G\"unther-Scharowsky-Str. 1 Bldg 24, 91058 Erlangen, Germany}
\affiliation{Institute of Optics, Information and Photonics, University Erlangen-Nuremberg, Staudtstr. 7/B2, 91058 Erlangen, Germany}

\begin{abstract}
The interaction of a cavity with an external field is symmetric under time reversal.
Thus, coupling to a resonator is most efficient when the incident light is the time reversed version of a free cavity decay, i.e. when it has a rising exponential shape matching the cavity lifetime. For light entering the cavity from only one side, the maximally achievable coupling efficiency is limited by the choice of the cavity mirrors' reflectivities. Such an empty-cavity experiment serves also as a model system for single-photon single-atom absorption dynamics.
We present experiments coupling exponentially rising pulses to a cavity system which allows for high coupling efficiencies.
The influence of the time constant of the rising exponential is investigated as well as the effect of  a finite pulse duration.
We demonstrate coupling 94\,\% of the incident TEM$_{00}$ mode into the resonator.
\end{abstract}

\maketitle

\section{Introduction}
Time-reversal symmetry is a fundamental concept in many branches of physics.
A recent overview of applications in optics can be found in~\cite{Leuchs2012}.
For example, time-reversal symmetry was exploited  to achieve efficient absorption of a microwave by an antenna~\cite{Lerosey2004} or strong focusing in the microwave domain~\cite{Lerosey2007} as well as in the optical domain~\cite{Vellekoop2010, Mudry2010}.
An early discussion about focusing onto an atom utilizing time-reversal arguments can be found in~\cite{Quabis2000}.
Furthermore, time-reversal techniques have found applications in light-matter interaction such as in the storage and retrieval of photons in atomic ensembles~\cite{Moiseev2001, Gorshkov2007, Novikova2007}.  Very recent experiments report on the experimental realization of time-reversed lasing~\cite{Wan2011}. Others utilize a time reversible scheme to optimize the quantum state transfer between two atoms trapped in distant cavities~\cite{Cirac1997, Ritter2012}.

Within the field of light-matter interaction, an experiment relying crucially on the application of time-reversal arguments is the absorption of a single photon by a single two-level system (TLS) in free space. Following the time-reversal argument given in~\cite{Sondermann2007, Stobinska2009}, the absorption probability can reach unity if the photon's shape is exactly the time reversed version of a spontaneously emitted photon and the momentum transfer to the atom can be neglected. For the temporal profile of the single-photon wavepacket this implies an exponentially rising envelope with a time constant matching the lifetime of the TLS. 

In recent experiments, time-reversal techniques were succesfully applied to demonstrate a high absorption probability of a single photon by an atomic ensemble~\cite{Zhang2012}. The benefit of pulse shaping in the case of single atoms and multi-photon pulses was demonstrated in~\cite{Aljunid2013}. However, experiments aiming at an efficient single-photon single-atom absorption process in free space remain challenging. Until now, absorption probabilities of 0.03\,\% have been reached in a three-level system~\cite{Piro2010}. 

To demonstrate the dynamics of such a free-space single-photon single-atom absorption experiment in a simple set-up, we make use of the analogy between a TLS and an empty optical resonator which was highlighted in earlier work~\cite{Heugel2009}.  Following this analogy, the energy stored inside the resonator is the equivalent to the probability of a single photon to be absorbed by the TLS, i.e. to the TLS to be found in the excited state. The field which is directly reflected by the cavity corresponds to the light transmitted by the atom. The field leaking from inside the cavity through the incoupling mirror is the counterpart to the light coherently scattered in direction of the transmitted beam while the cavity leakage through the end mirror is the counterpart  of the coherent scattering into all other directions. The cavity leakage through the incoupling mirror and the direct reflection (respective forward scattered and transmitted light) show destructive interference. In case of a perfectly reflecting 
end mirror ($R_2=1$) and an incident 
pulse with rising exponential shape matching the cavity lifetime these two fields have equal amplitudes and the  interference results in a zero net field amplitude during the pulse time-span. Until the end of the pulse there is no cavity reflection and 
the complete pulse energy is stored inside the resonator.
An imperfect end mirror ($R_2<1$) leads to unavoidable losses to the stored energy because the leakage through the end mirror cannot interfere with the incident light. This is equivalent to illumination of the atom from a limited solid angle smaller than $4\pi$ while the scattering occurs into the full solid angle. More precisely, the solid angle of illumination corresponds to the asymmetry $T_1/(T_1+T_2)$ of the cavity mirrors, with $T_1,\,T_2$ being the transmission coefficients of the incoupling and end mirror, respectively.

In this paper, we present the experimental implementation of such a coupling experiment, demonstrating the power of time-reversal symmetry in optics.

\section{Experimental set-up}

\begin{figure}
 \centering
 \includegraphics[width = .8\textwidth]{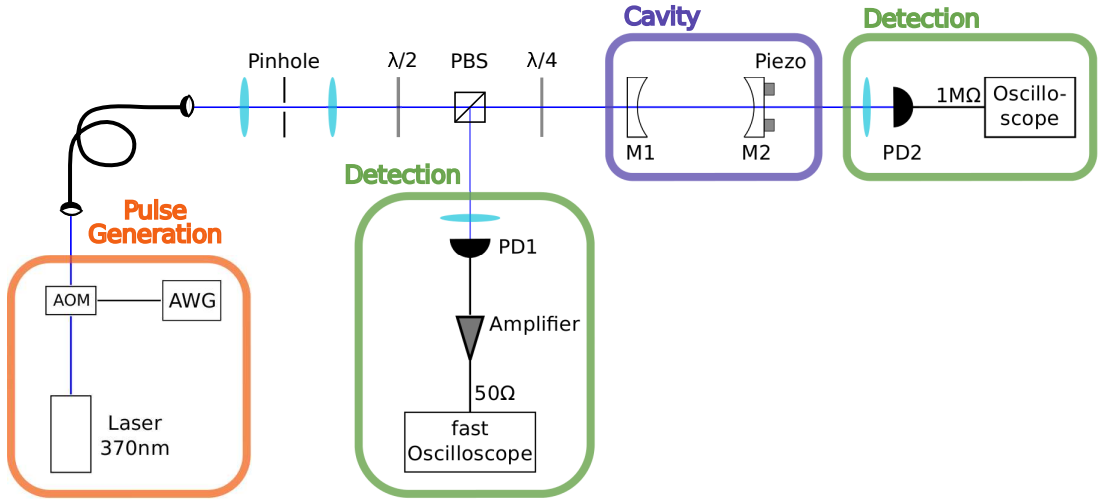}
 \caption{Schematic of the experimental set-up. The pulse shaping is implemented by an acousto-optic modulator (AOM) driven by an arbitrary-waveform generator (AWG). After transmission through an optical fibre, two lenses and a pinhole are used for spatial mode matching of the beam to the resonator. The detection of the cavity reflection is enabled by the combination of a polarizing beam-splitter (PBS) and a quarter-wave plate ($\lambda/4$). Cavity reflection and transmission are detected by the photodiodes PD1 and PD2, respectively.}
 \label{setup}
\end{figure}

Figure~\ref{setup} shows a schematics of the experimental set-up.
We use a frequency-doubled diode laser operating at a wavelength of 370\,nm. This is the transition wavelength of an Ytterbium ion which is our target system for the equivalent experiment~\cite{Maiwald2012}.
The desired pulses are cut out of the continuous-wave laser beam by means of an acousto-optic modulator (AOM) working at a center frequency of 400\,MHz. It is driven by an arbitrary-waveform generator (AWG) providing the radiofrequency signal with a suitably shaped envelope according to the desired pulse shape. The rise- and fall-time of the complete generation system amounts to 5\,ns and the repetition rate of the pulses is 200\,kHz. Details about the generation and characterization of rising exponential pulses can be found in~\cite{Golla2012}. Here we checked the pulse shapes with a fast photodiode (bandwidth 2\,GHz).

The pulsed beam is coupled to a single-mode optical fibre and sent to the resonator set-up. Two lenses and a filtering pinhole ensure the spatial mode matching of the beam to the cavity. The latter consists of two spherical mirrors at a distance of L=12.5\,cm, each having a focal length of 7.5\,cm. The reflectivity  of the incoupling mirror is determined by direct transmission measurements with a power meter. It amounts to $R_1=(97.96 \pm 0.07)\,\%$. According to manufacturer information, the scattering and absorption losses of the mirrors are two orders of magnitude smaller than $T_1$ and can thus be neglected in this work. The reflectivity of the end mirror is obtained from ring-down measurements of the asymmetric cavity revealing a lifetime of $\tau = (39.3 \pm 0.6)$\,ns. In combination with the knowledge of $R_1$ we deduce a reflectivity of $R_2 = (99.94 \pm 0.10)\,\%$. Since a reflectivity larger than one is unphysical, we set the upper bound of $R_2$ to 1 in the following. In order to tune the cavity 
length, the end mirror is mounted on a ring-shaped piezo actuator.
This cavity set-up corresponds to an atomic experiment where the incident light is focused from 97\,\% of the full solid angle. The lifetime of the cavity used here is comparable with atomic lifetimes.

By means of a polarizing beam-splitter followed by a quarter-wave plate the cavity reflection is separated from the incident beam. The reflected signal is detected by the photodiode PD1, amplified and finally recorded by a fast oscilloscope with a sampling rate of 25\,GS/s.  
The detection bandwidth is  limited by the bandwidth of the amplifier spanning the range from DC to 200\,MHz.

A second photodiode PD2 is installed to observe the cavity transmission. However, for an asymmetric cavity the leakage through the end mirror is very weak, resulting in a signal to noise ratio smaller than one when measuring with sufficient detection bandwidth for the coupling experiment. Thus, the complete dynamics of the system is extracted from the reflection signal PD1. PD2 in turn is used in 1\,M$\Omega$ coupling to determine the spatial mode matching of the incident beam to the cavity. For this purpose, the cavity length is scanned and transmission spectra of a continuous wave beam are taken. In the measurements presented here, the fraction $\eta$ of the incident light found in the TEM$_{00}$ mode lies between 94\,\% and 96.5\,\%.

\section{Data analysis method}		\label{SecAnalysis}

As mentioned above, the transmitted signal is too weak to be measured with the necessary temporal resolution due to the asymmetry of the cavity. Consequently we have to reconstruct the energy storage and the cavity leakage through the end mirror from the reflected signal making use of the experimentally determined mirror reflectivities.
In the following, we describe the  data analysis algorithm.

For each pulse shape we record two datasets: For the first one the cavity is tuned to resonance, for the second one the cavity is tuned far off resonance. In the latter case the pulse is reflected with hardly any interaction with the cavity. Thus, the corresponding dataset gives the incident pulse shape. In the following, the two datasets will be referred to as reflected intensity $I_{\text{R}}(t)$ and incident intensity $I_{\text{I}}(t)$, respectively.

In a first step we need to correct the data for imperfect spatial mode matching because light in higher-order spatial modes is not  resonant with the cavity in either measurement. This fraction of light will always be directly reflected and experiences no interference with the cavity leakage. Thus, we replace the incident intensity by $\eta I_{\text{I}}$ and the reflected intensity by $I_{\text{R}} - (1-\eta)I_{\text{I}}$ where $\eta$ is the intensity fraction of light in the TEM$_{00}$ mode. For simplicity, $I_{\text{I}}$ and $I_{\text{R}}$ will directly denote the mode matching corrected quantities in the following.

In the next step, we reconstruct the energy inside the cavity and the leakage through the end mirror as a function of time.
Let $E_{\text{I}}(t)$ denote the energy incident onto the cavity accumulated until time t, $E_{\text{R}}(t)$ the energy which has left the cavity in reflection until time t and $E_{\text{T}}(t)$ the corresponding quantity for the transmission. Then the energy stored inside the cavity can be written as

\begin{eqnarray}
	E_{\text{C}}(t) 	&= E_{\text{I}}(t) - E_{\text{R}}(t) - E_{\text{T}}(t)\\
			&= A \int_{-\infty}^t I_{\text{I}}(t') - I_{\text{R}}(t') - I_{\text{T}}(t')\,\text{d} t' 		\label{Eq:integral}
\end{eqnarray}

with A being the cavity area and $I_{\text{T}}$ the transmitted intensity. $I_{\text{I}}(t)$ and $I_{\text{R}}(t)$ are directly obtained from the measurement. The unknowns $E_{\text{C}}(t)$ and $I_{\text{T}}(t)$ are linked to each other by the electric field inside the cavity. Let $U_{\text{C}}(t)$ denote the amplitude of the field running in one direction. The standing wave inside the resonator is then built up by two counterpropagating waves each having the amplitude $U_{\text{C}}(t)$: 
\begin{equation}
	E_{\text{C}}(t) = AL\varepsilon_0 \vert U_{\text{C}}(t) \vert^2			\label{Eq:standingwave}
\end{equation}
The leakage through the end mirror is determined by its transmission coefficient $T_2$:
\begin{equation}
	I_{\text{T}}(t) = \frac{1}{2}c\varepsilon_0 \vert U_{\text{C}}(t) \vert^2 \cdot T_2			\label{Eq:transmission}
\end{equation}
Plugging \ref{Eq:standingwave} and \ref{Eq:transmission} into  \ref{Eq:integral} leads to an integral equation for $\vert U_{\text{C}}(t) \vert^2$:
\begin{equation}
	 \vert U_{\text{C}}(t) \vert^2 = \frac{1}{L\varepsilon_0}
	 \int_{-\infty}^{t} I_{\text{I}}(t') - I_{\text{R}}(t') - \frac{1}{2}c\varepsilon_0 T_2 \vert U_{\text{C}}(t') \vert^2\,\text{d} t'
\end{equation}
It is analytically solved yielding
\begin{equation}
	\vert U_{\text{C}}(t) \vert^2 = \frac{1}{L\varepsilon_0}\exp(-T_2t/t_{\text{rt}})
		\int_{-\infty}^{t} \left[ I_{\text{I}}(t')-I_{\text{R}}(t') \right]\exp(T_2t'/t_{\text{rt}})\,\text{d} t'
\end{equation}
with $t_{\text{rt}} = 2L/c$ being the round-trip time.

Inserting this result into  \ref{Eq:standingwave} we are able to reconstruct the energy stored inside the cavity from the measurement:

\begin{equation}		\label{Eq:Ec}
	E_{\text{C}}(t) = A\exp(-T_2t/t_{\text{rt}})
		\int_{-\infty}^{t} \left[ I_{\text{I}}(t')-I_{\text{R}}(t') \right]\exp(T_2t'/t_{\text{rt}})\,\text{d} t'
\end{equation}

Additionally we obtain the cavity transmission as

\begin{equation}
	E_{\text{T}}(t) = \frac{T_2}{t_{\text{rt}}}\int_{-\infty}^t E_{\text{C}}(t')\,\text{d} t' .
\end{equation}

Finally, all energies are normalized to the total energy $E_{\text{tot}}$ incident onto the cavity. In the implementation of this analysis, the lower bound of the integrals is set to the beginning of the measurement window which was chosen to cover at least the complete incident pulse and ten lifetimes of the cavity decay.

In the last step of the analysis process we search the maximum of $E_{\text{C}}$. This quantity will be referred to as energy coupling efficiency $E_{\text{C,max}}$ in the following. 

Since no assumptions on the incident pulse shape are made, the above method is applicable to any pulse shape.

\section{Results and discussion}

\begin{figure}
 \centering
 \includegraphics{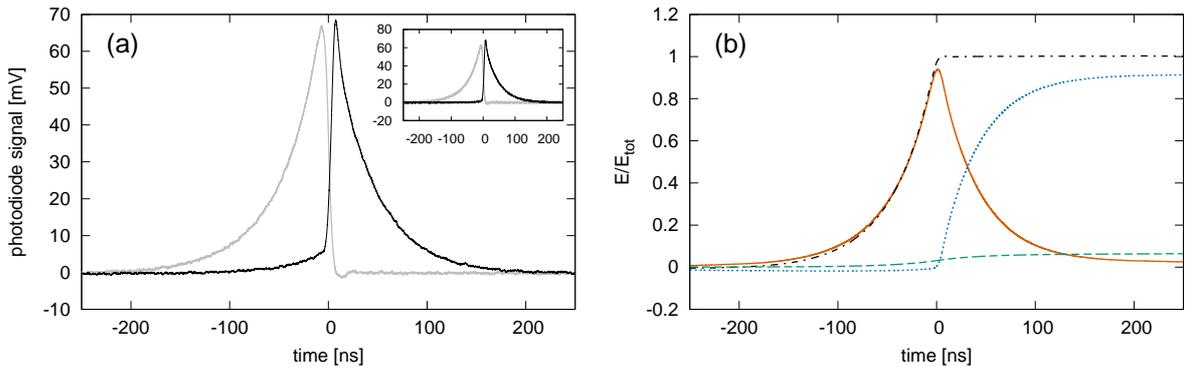}
 \caption{Intensity measurement and energy reconstruction for an exponentially rising pulse matching the cavity lifetime. (a) Measured intensity of the incident pulse (gray) and cavity reflection when the cavity is tuned to resonance (black). Inset: Same quantities after substraction of the non-mode-matched light. (b) Accumulated energy of incident pulse (dash-dotted), cavity reflection (dotted) and cavity transmission (dashed) as well as the energy stored inside the resonator (solid). The spatial mode matching amounts to $\eta=94.0\,\%$ in this measurement. For clarity, the plots are restricted to a time window shorter than the one used for measurement and analysis.}
 \label{SinglePulse}
\end{figure}

First, we discuss the analysis method outlined above using the example of an exponentially rising pulse with a time constant $\tau_{\text{P}}$ close to the cavity lifetime. All numbers given in this section refer to the fraction of the light which is spatially mode matched to the cavity. 
Figure~\ref{SinglePulse} shows the measured intensity traces as well as the reconstructed energies for a $\tau_{\text{P}} = 39.6$\,ns pulse. If the cavity is tuned into resonance (black line), the reflection is strongly suppressed until the end of the pulse. If plotting only the light in the TEM$_{00}$ mode (inset), hardly any rising of the photodiode signal can be observed until t=0, indicating that a huge amount of the energy is stored inside the cavity and only emitted after the end of the incident pulse. This is confirmed by the energy reconstruction displayed in figure~\ref{SinglePulse}~(b). The accumulated incident energy rises until the end of the pulse and remains then constant. The stored energy fraction follows closely nearly until the end of the pulse. The small difference can be explained by the finite fall time of the incident field after the radiofrequency signal driving the AOM was switched off. The 
reflected 
energy stays close to zero as long as the incident field is exponentially rising due to the interference of direct reflection and cavity leakage. When the incident pulse is switched off, the accumulated reflected energy amounts to only 1\,\% of the incident energy and 94\,\% are stored inside the cavity. For $t>0$ the cavity is decaying freely and most of the light is detected in reflection because of the asymmetry of the mirrors' reflectivities. In total, 7\,\% of the incident energy is transmitted through the cavity, partly already during the storage and partly during the decay. 

\begin{figure}
 \centering
 \includegraphics{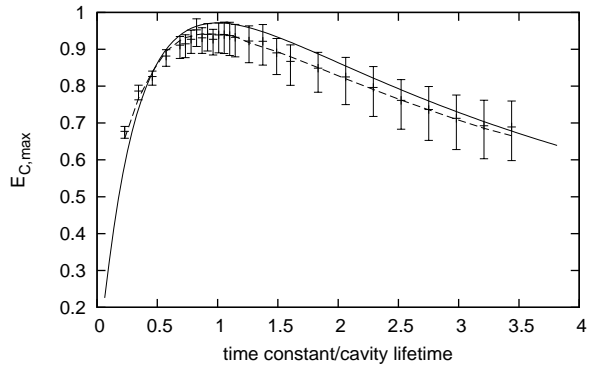}
 \caption{Energy fraction coupled to the resonator for exponentially rising pulses as a function of the time constant. Symbols: experimental values. Solid line: simulation for perfect pulse shapes. Dashed line: simulation using the measured incident pulse shapes as input signal. The spatial mode matching amounts to $\eta=94\,\%$ in this measurement.}
 \label{lifetimes}
\end{figure}

Next, we discuss the response of the cavity to exponentially rising pulses with different time constants. Figure~\ref{lifetimes} shows the experimental results for pulses each having a total duration of $10\tau_{\text{P}}$. 
 The error bars are based on the uncertainty of the mirror reflectivities. In agreement with the time-reversal argument, the energy coupling efficiency reaches a maximum when the time constant equals the independently measured cavity lifetime. The solid line reflects the theoretical prediction for perfect exponential pulses~\cite{Heugel2009}. The dashed line is the result of a simulation which takes the measured incident pulse shapes as input signal. The outcome of the latter simulation and the measured energy storage efficiencies agree very well quantitatively. The difference to the first simulation can be explained by the finite fall time of the AOM causing a smoothed edge at the end of the exponential pulse.
 
 For a pulse with time constant equal to the cavity lifetime, the effect of an imperfect temporal pulse shape can be described in a  particularly simple way: it reduces to a single number, namely the overlap of the incident field amplitude with the ideal one. In the measurement presented here, this field overlap amounts to $\eta_t = 98.6\,\%$. Following the reasoning of~\cite{Golla2012}, this implies an upper limit for the experimental energy coupling efficiency as $E_{\text{C,max}}\le\eta_t^2 = 97\,\%$.
 
 The same analysis could be performed for all other pulse shapes, bearing in mind that this simplified method is only valid for comparison with an ideal pulse, i.e. a perfect rising exponential pulse matching the cavity lifetime. Thus it doesn't reveal the difference to the simulation given by the solid line, but rather to its maximum, that is to say the maximally reachable coupling efficiency for a given pair of mirrors.
 
\begin{figure}
 \centering
 \includegraphics{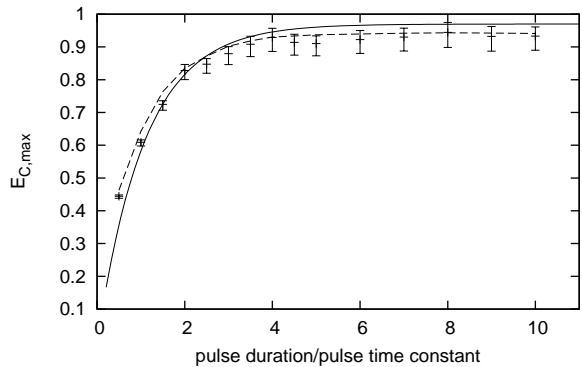}
 \caption{Energy fraction coupled to the resonator for an exponentially rising pulse as a function of the pulse duration. The time constant of the incident pulse is fixed to $\tau_{\text{P}} = 36$\,ns. Symbols: experimental values. Solid line: simulation for perfect pulse shapes. Dashed line: simulation using the measured incident pulse shapes as input signal. The spatial mode matching amounts to $\eta=96.5\,\%$ in this measurement.}
 \label{durations}
\end{figure}

According to the time-reversal argument~\cite{Heugel2009}, the perfectly matched pulse has to start at $t=-\infty$ with an infinitely small intensity. However, in experiments pulse generation is limited to a finite time window which we refer to as pulse duration. The influence of finite pulse durations on the energy coupling efficiency is displayed in figure~\ref{durations}. For this plot, we created exponentially rising pulses each having a time constant of $\tau_{\text{P}} = 36$\,ns with durations varying between $0.5\tau_{\text{P}}$ and $10\tau_{\text{P}}$. 
For long pulses, the energy coupling efficiency shows an asymptotic behaviour. The asymptote is approached in good approximation for pulse durations of four times the pulse time constant or longer. Thus experiments as they underly figure~\ref{lifetimes} with pulses of duration $10\tau_{\text{P}}$  can be considered as experiments with practically infinitely long pulses.

\begin{figure}
 \centering
 \includegraphics{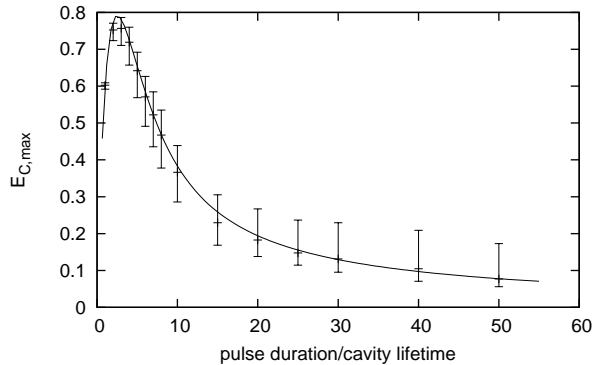}
 \caption{Energy fraction coupled to the resonator for rectangular pulse shapes. Symbols: experimental values. Solid line: simulation for perfect pulse shapes.  The spatial mode matching amounts to $\eta=96.5\,\%$ in this measurement.}
 \label{rects}
\end{figure}

Finally we measured the energy incoupling efficiency for rectangular pulses with different durations (figure~\ref{rects}). In this case, the deviation of the incident pulse from a perfect rectangular shape has hardly any influence. The measured energy coupling efficiency shows very good agreement with the theoretical prediction.

\section{Conclusion}
We implemented a cavity set-up with an asymmetry of $T_1/(T_1+T_2) = 97\,\%$ to simulate the absorption dynamics of a single photon by a single atom. Experiments with exponentially rising incident pulses demonstrate the power of time-reversal arguments in optics: by choosing a time constant matching the cavity lifetime, we achieved an energy coupling efficiency of $E_{\text{C,max}}=94\,\%$. As defined in section~\ref{SecAnalysis}, this value refers to the fraction of the light which is spatially mode matched to the cavity.

While for an ideal system the complete pulse energy can be coupled to the resonator, any real implementation is subject to several restrictions as summarized in table~\ref{Tab:Restrictions}. A conceptual limitation is given by the choice of the mirror reflectivities. In an experiment as performed here, the maximally attainable coupling efficiency is exactly the asymmetry of the mirrors' reflectivities~\cite{Heugel2009}. In an atomic absorption experiment, a similar constraint is given by the geometry of the set-up limiting the solid angle used for focusing onto the atom. The spatial beam shaping and the temporal pulse forming are technical restrictions for both kinds of experiments. Taking into account also light in higher-order spatial modes, the experimentally observed fraction of the total incident energy stored inside the resonator amounts to 88\,\%. Considering 
all these constraints to our implementation, the measured energy coupling agrees very well with the theoretical prediction.
\begin{table}
\caption{\label{Tab:Restrictions}Summary of the restrictions our experiment is subject to and deduced energy coupling efficiencies. The experimentally achievable values refer to the parameters used in this experiment.}
\begin{tabular}{p{5mm}p{50mm}c}
      \hline
      \multicolumn{3}{l}{\textbf{Experimental restrictions}}\\
      \hline
      & asymmetry $T_1/(T_1+T_2)$ &    97\,\%\\
      & spatial mode matching $\eta$&    94\,\%\\
      & temporal mode matching $\eta_t^2$&       97\,\%\\
      \hline      
      \multicolumn{3}{l}{\textbf{Coupling efficiency}}    \\
      \hline
      Ideally				&&	100\,\%\\
      \multicolumn{2}{l}{For spatially mode matched fraction}&\\
      & experimentally achievable	&	94\,\%\\
      & measured value			& 	$94^{+3}_{-5}$\,\%\\
      \multicolumn{2}{l}{For total incident power}&\\
      & experimentally achievable	&	88\,\%\\
      & measured value			& 	$88^{+3}_{-5}$\,\%\\
      \hline
\end{tabular}
\end{table}

\section{Acknowledgements}

M.S. acknowledges financial support from the Deutsche
Forschungsgemeinschaft (DFG). G.L. wants to thank the
German Federal Ministry of Education and Research
(BMBF) for financial support in the framework of the
joint research project \enquote{QuORep}.

\bibliographystyle{iopart-num}

\providecommand{\newblock}{}

\end{document}